\documentclass[12pt]{article}
\pdfoutput=1
\usepackage[a4paper]{geometry}
\usepackage{amsmath,amssymb,amsfonts,latexsym}
\usepackage{graphicx}  
\usepackage[colorlinks=true, pdfstartview=FitV, linkcolor=blue, citecolor=blue, urlcolor=blue]{hyperref}

\newcommand{\bra}{\left\langle} 
\newcommand{\ket}{\right\rangle}
\newcommand{\ep}{\qquad {\vrule height 10pt width 8pt depth 0pt}}

\newcommand {\bR}{{\mathbb R}}
\newcommand {\bN}{{\mathbb N}}

\newcommand {\bZ}{{\mathbb Z}}
\newcommand {\bI}{{\mathbb I}}

\newcommand {\bB}{{\mathbb B}}

\newcommand {\bK}{{\mathbb K}}

\newcommand {\bQ}{{\mathbb Q}}

\newcommand {\cC}{{\cal C}}

\newcommand {\cO}{{\cal O}}
\newcommand {\cS}{{\cal S}}

\newtheorem{theorem}{Theorem} [section]
\newtheorem{lemma}[theorem]{Lemma}
\newtheorem{propo}[theorem]{Proposition}

\newtheorem{corollary}[theorem]{Corollary}

\newtheorem {remark}[theorem]{Remark}
\newtheorem {remarks}[theorem]{Remarks}

\begin{document}
\title{Stability of the electron cyclotron resonance}
\author{
Joachim Asch  \thanks{Aix Marseille Universit\'e, CNRS, CPT UMR 7332, F--13288 Marseille cedex 9, France, e-mail : asch@cpt.univ-mrs.fr}\ 
\thanks{Universit\'e de Toulon, CNRS, CPT UMR 7332,  B.P. 20132, F--83957 La Garde, France}\ 
Olivier Bourget\thanks{
Facultad de Matem\'aticas,Pontificia Universidad Cat\'olica de Chile, Av. Vicu\~{n}a Mackenna 4860, C.P. 690 44 11, Macul, Santiago, Chile},
C\'edric Meresse\thanks{Aix Marseille Universit\'e, CNRS, CPT UMR 7332, F--13288 Marseille cedex 9, France}
}
\date{11.07.2015}
\maketitle

\abstract{We consider the magnetic AC Stark effect for the quantum dynamics of a single particle in the plane under the influence of an oscillating homogeneous electric and a constant perpendicular magnetic field. We prove that the electron cyclotron resonance is insensitive to impurity potentials.}

\section{Introduction}
 For $T>0$ let $E\in C^{0}(\bR,\bR^{2})$ be a $T$ periodic function  and $V$ be a $T$ periodic multiplication operator by a real valued bounded function  in $L^{2}(\bR^{2},dq)$.
The time dependent family of Hamiltonians  under investigation is
\[H(t)=H_{0}(t)+ V(t)\qquad(t\in\bR)\]
with
\[H_{0}(t)=H_{La} - \langle E(t),q\rangle\hbox{\ \em\it and \ } H_{La}=\frac{1}{2}\left(-i\nabla-\frac{q^{\perp}}{2}\right)^{2},\]
where $a^{\perp}$ denotes the direct perpendicular of a vector $a$.
$H_{La}, H_0(t), H(t)$ are essentially selfadjoint on the  Schwartz space $\cS(\bR^{2})$.

$H(\cdot)$ describes the dynamics of a quantum particle of mass $m$ and charge $e$ for  a magnetic field of strength $B$ with    $eB>0$, in units of magnetic length $\sqrt{\frac{\hbar}{ e B}}$,  gyration time $\frac{m}{ e B}$, and energy  $\frac{\hbar e B}{m}$.  $E$ defines the strength of the  electric field and $V$  the  impurity potential.

Electron cyclotron resonance means the growth of the system's kinetic energy if the frequency of the electric field is in resonance with the cyclotron frequency defined by the magnetic field and the particle:  the spectrum of $H_{La}$ is pure point and equals $\bN+\frac{1}{2}$;   thus $e^{-i 2\pi H_{La}}=e^{-i \pi}\bI$,  all orbits generated by $H_{La}$ are periodic with cyclotron frequency equal to $1$;  in case of resonance, i.e. if $T$ is an integer multiple of $2\pi$, the Floquet operator of $H_0$  turns out to be a product of phase space translation operators, see (\ref{eq:Floquet}), which may accelerate the system. 

We shall prove that this acceleration and also the asymptotic velocity is insensitive to the impurity $V$ in a large class of potentials provided that the force field created by $V$ vanishes at spatial infinity. 

We state this more precisely. Denote the Fr\'echet space of smooth bounded functions whose derivatives are bounded 

\[S(1):=\left\lbrace f\in C^{\infty}(\bR^{2};\bR); \sup_{q\in\bR^2}\vert\partial^{\alpha}f(q)\vert<\infty, \quad \alpha\in\bN_0^2\right\rbrace.\]

We always make the assumption

\medskip
\noindent {\bf (A)}: {\it Let $T>0$. Suppose that $E\in C^{0}(\bR;\bR^{2})$ and $V\in C^{0}(\bR; S(1))$  are $T$ periodic functions.}

\medskip

The family $H(\cdot)$ then defines a unitary propagator $U$, see Theorem \ref{thm:propagator}, and is $T$ periodic so  one knows from Floquet's Theorem that $U(t+T,t_0+T)=U(t,t_0)$ for all $t,t_0$. Denote  $U(T):=U(T,0)$ its Floquet operator,
\[R(t):=\left(\begin{matrix}\cos t&-\sin t\\ \sin t&\cos t\end{matrix} \right)\]
 the rotation matrix of angle $t$, $q$ the (vector) multiplication operator by the coordinate, 
  and the  operator of asymptotic velocity
\[v_{asy}:=s-\lim_{n\to\infty}\frac{1}{nT}\left(U^{\ast}(nT)qU(nT)-q\right)\]
on its maximal domain.

We summarize our main results in two theorems. If the force $\nabla V$ vanishes at spatial infinity we have

\begin{theorem}\label{thm:main} Assume {\bf (A)} and suppose
$\vert\nabla V(t,q)\vert\to_{\vert q\vert\to\infty}0$, uniformly in $t$. 

\medskip\noindent If $T\in2\pi\bN$ and $\left\vert \int_{0}^{T}E(t)\ dt\right\vert+\left\vert\int_{0}^{T}R(t)E(t)\ dt\right\vert >0$ then

\begin{enumerate}
\item the spectrum of $U(T)$ is purely absolutely continuous except possibly for a finite number of eigenvalues;
\item $v_{asy}$ has discrete spectrum independent of $V$ and :
\[v_{asy}=\frac{1}{T}\int_{0}^{T}\left(R(t)E^{\perp}(t)-E^{\perp}(t)dt\,\right) P_{ac}(U(T))\]
$P_{ac}$ denoting the projector on the absolutely continuous subspace;

\item for $\psi$ in the  domain of $H_{La}$  it holds
\[\lim_{n\to\infty}\frac{1}{(nT)^2}\left\langle U(nT)\psi,H_{La}U(nT)\psi\right\rangle=\rho\Vert P_{ac}(U(T))\psi\Vert^2 \]
with the $V$-independent rate
\[\rho=\frac{1}{2}\left\vert\frac{1}{T}\int_{0}^{T}R(t)E(t)\ dt\right\vert^{2}.\]
\end{enumerate}

\medskip \noindent If  $T\notin2\pi\bN$  it holds
 \begin{enumerate}
 \item  if $\int_{0}^{T}E(t)\ dt\neq0$ 
\begin{enumerate}
\item the spectrum of $U(T)$ is purely absolutely continuous except possibly for a finite number of eigenvalues;
\item $v_{asy}$ has discrete spectrum independent of $V$:
\[v_{asy}=\left(-\frac{1}{T}\int_{0}^{T}E^{\perp}(t) dt\right)P_{ac}(U(T));\]
\end{enumerate}
 \item  if  $\int_{0}^{T}E(t)\ dt=0$ then $v_{asy}=0$.
 \end{enumerate}  
\end{theorem}

Remark that we adopt the convention to count eigenvalues with their multiplicities;  the absolutely continuous subspace in Theorem \ref{thm:main}  is of finite codimension.

\medskip
If $\nabla V$ is small with respect to the external field $E$ or if $V$ is local in space we have

\begin{theorem}\label{coro:misc} Assume {\bf (A)}.
\begin{enumerate}
\item \label{item:one} Suppose $\left\vert \int_{0}^{T}E(t)\ dt\right\vert >0$  and $\sup_{t,q}\vert\nabla V(t,q)\vert< \left\vert\frac{1}{T}\int_0^T E\right\vert$ or \\$\left\vert\int_{0}^{T}R(t)E(t)\ dt\right\vert >0$ and $\sup_{t,q}\vert\nabla V(t,q)\vert< \left\vert\frac{1}{T}\int_0^T RE\right\vert$ then the spectrum of $U(T)$ is purely absolutely continuous.
\item \label{item:local} Suppose $V(t,q)\to_{q\to\infty}0$ uniformly in $t$. If $T\in2\pi\bQ$ and $\int_{0}^{T}E=0=\int_{0}^{T}R(s)E(s)\ ds$ then the spectrum of $U(T)$ is pure point.

\end{enumerate}\end{theorem}

Several remarks are in order.
\begin{remarks}\label{remarks:comment}
\begin{enumerate}
\item Our results apply in particular to the case of constant $E$ which describes the static Quantum Hall Effect. We refer to \cite{g}  for a review of known mathematical results. These concern mainly the small $E$ limit of vanishing exterior electric field and potentials $V$ which are random or periodic.
\item In Theorem \ref{thm:main} no decay of  $V$ is assumed, on the other hand, due to the decay assumption on $\nabla V$ we do not say anything for random or periodic $V$. 
\item In Theorem \ref{coro:misc}.\ref{item:one} smallness of  $\nabla V$ with respect to $E$  implies that  the spectrum of $U(T)$ is purely absolutely continuous.  For fixed $V$ the condition is not fulfilled when $E$ vanishes. We have no explicit information on asymptotic acceleration or velocity in this case. 
\item $v_{asy}$ may be considered as proportional to the time averaged Hall-current. If  $E$ is constant we have in particular $v_{asy}=-E^\perp P_{ac}(U(T))$. Remark that the time dependency of the electric field can suppress $v_{asy}$ in the resonant case; as an example : if $E(t):=E_0+R(-t)E_0$, for an $E_0\in\bR^2$, then $v_{asy}=0$. 
\item It suffices to assume $\int_{0}^{T}E\neq 0$ and $\bra \int_{0}^{T}E^{\perp},\nabla V(t,q)\ket\to_{\vert q\vert\to\infty}0$ uniformly in $t$ to conclude that the absolutely continuous subspace is of finite codimension. This is a corollary of the proof of Theorem \ref{thm:main}. The condition is, however, not sufficient so deduce information on $v_{asy}$ in the direction of $\int_0^T E$.
\item \label{remarks:stability}In Theorem \ref{coro:misc}.\ref{item:local}  we prove that the spectrum of $U(T)$ is pure point in the resonant case. So local potentials are not likely to accelerate the system. We guess that this remains true in the non resonant case; however, no method to prove this seems to be available. The problem of stability under non-resonant  time periodic local perturbations  was raised in \cite{ev}. Progress on this difficult problem has  recently been achieved  for perturbations of the harmonic oscillator in one dimension \cite{w,gt}. \end{enumerate}
\end{remarks}

In \cite{gy} the electric AC Stark effect was discussed for perturbations of the isotropic harmonic oscillator. They proved for a sinusoidal electric field that the spectrum of the Floquet hamiltonian is  purely absolutely continuous if $\nabla V$ is small.

\cite{bf} observed  \ref{coro:misc}.\ref{item:one} for the  case $\vert\nabla V\vert< \left\vert\frac{1}{T}\int_0^T E\right\vert$.

The eigenvalues of the asymptotic velocities were examined in \cite{ak} for $-\frac{\hbar^2}{2}\Delta+V$ and space-periodic $V$ in the Floquet-Bloch decomposition.

We shall discuss  general properties of the propagator in section \ref{sec:propagator}. Then in section \ref{sec:compactness} we establish Theorem \ref{thm:compactness},   a dynamical compactness result which is our main technical contribution. We prove a Mourre estimate in Theorem \ref{thm:fulldynamics}; this allows us to apply a unitary Mourre Theorem which implies the spectral results. We  finish the paper with the proof of Theorems \ref{thm:main}, \ref{coro:misc}.

\section{The propagator}\label{sec:propagator}
$H_{0}(\cdot)$ is quadratic in the canonical operators; as in the case of the Stark effect \cite{ah, ky} its propagator can be determined explicitly. We present it here  in terms of center and velocity operators. Although this is folklore we provide an explicit proof. See  \cite{fo} for the metaplectic representation in general. 

In the following, if not otherwise specified, operators and identities are to be understood as acting on Schwartz-space $\cS(\bR^{2})$. Denote $D:=-i\nabla$. The (vector-) operators of center $c$ and velocity $v$ are
\[v=(v_{1},v_{2}):=D-\frac{1}{2}q^{\perp};\qquad c=(c_{1},c_{2}):=\frac{1}{2}q-D^{\perp}.\]
Relevant commutation relations are
\begin{eqnarray}
 [v_1,v_2]=i\,\, ,\quad[c_2,c_1]=i\,\, ,\quad[c_j,v_k]=0 \quad\forall j,k\nonumber\\
\quad\lbrack c, H_{La}\rbrack=0 ,  \lbrack v^{\perp},H_{La}\rbrack=i v.\phantom{go further to the left}\hfill\label{eq:commutation relations}
\end{eqnarray}
To relate to the usual position and momentum  operators note that
\[q=c+v^{\perp}, \quad D=\frac{1}{2}\left(c^{\perp}+v\right).\]
While the operator of kinetic energy is $H_{La}=\frac{1}{2}v^{2}$ the name center operator is motivated by the relation
\begin{equation}e^{i t H_{La}}q e^{-itH_{La}}=c+R(-t)v^{\perp}\label{eq:qla}\end{equation}
which follows from (\ref{eq:commutation relations}).

Denote for $a\in\bR^{2}$ and a vector operator  $w$: 
\[\bra a,w\ket:=a_{1}w_{1}+a_{2}w_{2}.\]

This an abuse of notation as we also use $\langle\cdot,\cdot\rangle$ for the scalar product in $L^2(\bR^2)$ in the statement of Theorem \ref{thm:main}.

Note that the domain of $H(t)$ is time dependent in general so some care is needed to assure the existence of its propagator. We have
\begin{theorem}\label{thm:propagator}
Assume {\bf (A)}. For $t\in\bR$ consider the operators
\[H_{La}:=\frac{1}{2}\left( D-\frac{1}{2}q^{\perp}\right)^{2},\quad H_{0}(t):=H_{La}-\bra E(t),q\ket, \quad H(t)=H_{0}(t)+V(t)\]
which are all essentially selfadjoint on $\cS(\bR^{2})$.  We then have:

\begin{enumerate}
\item 

the family $H(\cdot)$ generates a unitary propagator $U$ which leaves $\cS(\bR^{2})$ invariant. Furthermore for $\psi\in\cS(\bR^{2})$, $t\mapsto U(t,t_{0})\psi$ is the solution of 
\[i\partial_{t}\psi(t)=H(t)\psi(t), \psi(t_{0})=\psi;\]

\item

$S$ defined by
\[S(t,t_{0}):=e^{i\bra\int_{t_{0}}^{t}E,c\ket}e^{i\bra\int_{t_{0}}^{t}RE,v^{\perp}\ket}e^{-i\varphi(t,t_{0})}\]
with
\[\varphi(t,t_{0}):=\int_{t_{0}}^{t}\left(\frac{1}{2}\bra E(s),\int_{0}^{s}E^\perp\ket-\frac{1}{2}\bra R(s)E(s),\int_{0}^{s}RE^\perp\ket\right)\ ds, \]

is a unitary propagator;

\item 

for the propagator $U_{0}$ generated by $H_0(\cdot)$   it holds: 
\begin{equation}U_{0}(t,t_{0})=e^{-i t H_{La}}S(t,t_{0})e^{i t_{0}H_{La}};\end{equation}

\item \label{4}

$U$ is given by  
\begin{equation}U(t,t_{0})=U_{0}(t,0)\Omega(t,t_{0})U_{0}(0,t_{0})\label{eq:freepropagator}\end{equation}

where $\Omega$ is the propagator generated by the family of bounded operators

\[V\left(t,x(t,q,D)\right)\]
which is the Weyl quantization of the symbol $V(t,x(t,q,p))$ 
with 
\begin{equation}\label{eq:xofqp}
x(t,q,p):=\frac{1}{2}q-p^\perp-\int_{0}^{t}E^\perp+R(-t)\left(\frac{1}{2}q+p^\perp+\int_{0}^{t}RE^\perp\right)
\end{equation}

for $q,p\in\bR^2$. In particular it holds for $T=n2\pi, \quad n\in\bN$
\begin{eqnarray}\label{eq:Floquet}
U_{0}(T)&:=&U_{0}(T,0)=e^{i \alpha_{n}}e^{i\bra\int_{0}^{T}E,c\ket}e^{i\bra\int_{0}^{T}RE,v^{\perp}\ket} \\
&=&e^{i\beta_{n}}e^{\frac{i}{2}\bra\int_{0}^{T}(\bI+R)E,q\ket}e^{i\bra\int_{0}^{T}(\bI-R)E^{\perp},D\ket}\nonumber
\end{eqnarray}
with $i\alpha_{n}:=-i\pi n-i\varphi(T)$,\\ $i\beta_{n}:=-i\pi n-i\varphi(T)+\frac{i}{4}\bra\int_{0}^{T}(\bI-R)E^{\perp},\int_{0}^{T}(\bI+R)E\ket$, \\
and $\bI$ the $2\times2$ identity matrix.

\end{enumerate}
\end{theorem}

\medskip
{\bf Proof.} We use the commutation relations (\ref{eq:commutation relations}). Note that for observables $w_{\ell}$ which are  linear polynomials in   $D_{j},q_{k}$ it holds
\begin{equation}\label{eq:bch0}e^{-i(w_{1}+w_{2})}=e^{-iw_{1}}e^{-iw_{2}}e^{\frac{1}{2}\lbrack w_{1},w_{2}\rbrack}
\end{equation}
which for the case $\lbrack w_{1},w_{2}\rbrack=i$ and $a\in C^{1}(\bR;\bR^{2})$ implies  for $\psi\in\cS(\bR^{2})$
\begin{equation}\label{eq:bch}
i\partial_{t}e^{-i\bra a,w\ket}\psi=\left(\bra\dot{a},w\ket+\frac{1}{2}\bra\dot{a},a^{\perp}\ket\right)e^{-i\bra a,w\ket}\psi.
\end{equation}

Furthermore  $e^{itH_{La}}v^{\perp}e^{-itH_{La}}=R(-t)v^{\perp}$ because  the derivatives of both functions coincide as well as their values at $t=0$. This also proves (\ref{eq:qla}).

To apply formula (\ref{eq:bch}) we write
\[S=e^{-i \bra a,c\ket}e^{-i\bra b, v^{\perp}\ket}e^{-i\varphi}\]
with $a(t):=-\int_{t_{0}}^{t}E$, $b(t):=-\int_{t_{0}}^{t}RE$ and conclude for $\psi\in\cS(\bR^{2})$:
\begin{eqnarray*}
i(\partial_{t}S) S^{-1}\psi&=&\left(\bra \dot{a},c\ket+\bra \dot{b},v^{\perp}\ket-\frac{1}{2}\bra\dot{a},a^{\perp}\ket+\frac{1}{2}\bra\dot{b},b^{\perp}\ket+\dot{\varphi}\right)\psi\\
&=&\left(-\bra E,c\ket-\bra RE,v^{\perp}\ket\right)\psi
\end{eqnarray*}

by definition of $\varphi$.

Also:
\begin{eqnarray*}
i\partial_{t}e^{-it H_{La}}S(t,0)\psi&=&
\left(H_{La}+e^{-itH_{La}}i\partial_{t}SS^{-1}(t)e^{itH_{La}}\right)e^{-itH_{La}}S(t,0)\psi\\
&=&\left(H_{La}-\bra E(t),c\ket-\bra R(t)E(t),R(t)v^{\perp}\ket\right)e^{-itH_{La}}S(t,0)\psi\\
&=&H_{0}(t) e^{-itH_{La}}S(t,0)\psi.
\end{eqnarray*}
It follows that $U_{0}(t,0)=e^{-itH_{La}}S(t,0)$;  (\ref{eq:freepropagator}) results from $U_{0}(t,t_{0})=U_{0}(t,0)(U_{0}(t_{0},0))^{-1}$.

The formula for $U_{0}(T)$ in terms of $q,D$ follows from the identity
\[\bra a,c\ket+\bra b,v^{\perp}\ket=\bra\frac{a+b}{2},q\ket+\bra(a-b)^{\perp},D\ket\]
and (\ref{eq:bch0}).

Now define $\Omega(t,0):=U^{\ast}_{0}(t,0)U(t,0)$ then
\[i\partial_{t}\Omega(t,0)\psi=U_{0}^{\ast}(t,0)V(t)U_{0}(t,0)\Omega(t,0)\psi;\]
to prove the equality $U^{\ast}_{0}VU_{0}(t)=V\left(t,x(t,q,D)\right)$
note that for $a\in\bR^{2}$  it follows from (\ref{eq:commutation relations}):
\[e^{i\bra a, v^{\perp}\ket}v^{\perp}e^{-i\bra a, v^{\perp}\ket}=v^{\perp}-a^{\perp},\qquad
e^{i\bra a, c\ket}ce^{-i\bra a, c\ket}=c+a^{\perp}.\]
Together with (\ref{eq:qla}) this yields
\begin{eqnarray}
&&U^\ast_0(t,0)qU_0(t,0)=c-\int_{0}^{t}E^\perp+R(-t)\left(v^\perp+\int_{0}^{t}RE^\perp\right)\label{eq:qoftfree}\\ 
&&=x(t,q,D)=
\frac{1}{2}q-D^\perp-\int_{0}^{t}E^\perp+R(-t)\left(\frac{1}{2}q+D^\perp+\int_{0}^{t}RE^\perp\right)\nonumber.
\end{eqnarray}
The  equality $U^{\ast}_{0}VU_{0}(t)=V\left(t,x(t,q,D)\right)$ now follows from Egorov's Theorem because $H_0$ is a quadratic polynomial in $q$ and $D$, cf \cite{fo}.
By the Calderon Vaillancourt Theorem the operator $V\left(t,x(t,q,D)\right)$ is bounded thus the propagator $\Omega$ is well defined. The invariance of Schwartz space now follows from \\ $i\partial_t\psi(t)=V\left(t,x(t,q,D)\right)\psi(t)$ and the fact that the pseudodifferential operator $V\left(t,x(t,q,D)\right)$ leaves $\cS(\bR^2)$ invariant.
 \ep

Note the free evolution of the observables $c, v^\perp, H_{La}$ :

\begin{corollary}\label{coro:freepropagator} For $\psi\in\cS(\bR^2)$ and $t\in\bR$ we have
\begin{eqnarray}U^\ast_0(t,0)cU_0(t,0)\psi=c\psi-\int_{0}^{t}E^{\perp}\psi\label{eq:coftfree}\\ U^\ast_0(t,0)v^\perp U_0(t,0)\psi=R(-t)\left(v^{\perp}\psi+\int_{0}^{t}RE^{\perp}\psi\right)\label{eq:voftfree}
\end{eqnarray}
\[U^\ast_0(t,0)H_{La}U_0(t,0)\psi=\frac{1}{2}\left(v^{\perp}+\int_{0}^{t}RE^{\perp}\right)^{2}\psi.\]
\end{corollary}

By Floquet's Theorem it holds
\[U(t)=M(t)\left(U(T)\right)^{\frac{t}{T}}\]
for a unitary  $T$ -- periodic family $M$. Eigenvectors of $U(T)$ give rise to periodic orbits whereas for $\psi$ in its absolutely continuous subspace the function  $n\mapsto\langle \psi,U(nT)\psi\rangle$ decays. 
Define
\begin{equation}a_{c}(t):=-\int_{0}^{t}E^{\perp},\qquad a_{v}(t):=\int_{0}^{t}RE^{\perp}\label{eq:acav}\end{equation}

By (\ref{eq:Floquet}), for $T\in2\pi\bN$, the operator $U_{0}(T)=e^{i \alpha_{n}}e^{i\bra a_c^\perp,c\ket}e^{-i\bra a_{v}^\perp,v^{\perp}\ket} $ is, except for a phase,  the product of the two commuting phase space translation operators; its spectrum is absolutely continuous if $a_{c}(T)$ or $a_{v}(T)$ is non zero.

In order to prove spectral properties for $U(T)$  we shall use the following result of unitary Mourre theory which was proven  in \cite{abcf}, Theorem 3.3. Consult also \cite{abc}, Theorem 2.3, for an optimal version.

\begin{theorem}\label{thm:abcf}
Let $U$ be unitary and $A$ selfadjoint in a separable Hilbert space such that 
\[(U^\ast A U - A) \quad \hbox{ and }\quad \left\lbrack (U^\ast A U-A), A\right\rbrack\]
are densely defined on an $U$-invariant core of $A$ and can be extended as bounded operators. Suppose that
\[(U^\ast A U - A)\ge c \bI + K\]
for a $c>0$ and $K$ a compact operator. Then the spectrum of $U$ is absolutely continuous except possibly for a finite number of eigenvalues of finite multiplicity. If $K=0$ the spectrum is purely absolutely continuous.
\end{theorem}

Define the Mourre operators 
\begin{equation}A_{c}:=\bra a_{c}(T),c\ket,\qquad A_{v}:=\bra a_{v}(T),v^{\perp}\ket.\label{eq:aoperators}\end{equation}

By the explicit evolution of observables in the free case we have

\begin{propo}\label{propo:freeMourre} Let $T>0$, suppose $E\in C^0(\bR, \bR^2)$ and $T$ periodic. Denote  $\sigma$ the spectrum of $U_{0}(T)$. It holds:
\begin{equation}\left(U_{0}^{\ast}(T)A_{c}U_{0}(T)-A_{c}\right)=\vert a_{c}(T)\vert^{2}.\label{eq:Mourrecfree}\end{equation}
\begin{enumerate}
\item If $a_{c}(T)\neq0$ then 
 $\sigma$ is  purely absolutely continuous.
\item If $a_{c}(T)=0$ then 
\begin{enumerate}
\item {\it if } $T\notin 2\pi\bN$ {\it then } $\sigma$ is pure point
\item { \it if } $T\in 2\pi\bN$ {\it then } 
\begin{equation}\left(U_{0}^{\ast}(T)A_{v}U_{0}(T)-A_{v}\right)=\vert a_{v}(T)\vert^{2},\label{eq:Mourrevfree}\end{equation}
{\it if } $a_{v}(T)\neq0${ \it then }$\sigma$ is purely absolutely continuous, {\it if} $a_{v}(T)=0$ then $\sigma$ is pure point.
\end{enumerate}
\end{enumerate}
\end{propo}

{\bf Proof.} The identities (\ref{eq:Mourrecfree}), (\ref{eq:Mourrevfree})  follow from Corollary \ref{coro:freepropagator}. The second order commutators, which are densely defined on $\cS(\bR^2)$,  vanish:
 \[\left\lbrack\left( U_{0}^{\ast}(T)A_{c}U_{0}(T)-A_{c}\right), A_{c}\right\rbrack=\lbrack\left( U_{0}^{\ast}(T)A_{v}U_{0}(T)-A_{v}\right), A_{v}\rbrack=0.\]
It follows from Theorem \ref{thm:abcf} that the spectrum of $U_{0}(T)$ is purely absolutely continuous if $a_{c}(T)\neq0$ for any $T$ or $a_{v}(T)\neq0$ for $T\in2\pi\bN$.
 
 Concerning the remaining cases, remark that for $\psi\in\cS(\bR^{2})$ and $t\in\bR$ it holds
\[\sup_{t\in\bR}\Vert \left(\bI+c^{2}+v^{2}\right)U_{0}(t)\psi\Vert=\sup_{t\in\bR}\Vert\left(\bI+\left(c+a_{c}(t)\right)^{2}+\left(v^{\perp}+a_{v}(t)\right)^{2}\right)\psi\Vert.\]
This supremum is finite iff $a_{c}$ and $a_{v}$ are bounded functions of time.
On the other hand $\left(1+c^{2}+v^{2}\right)^{-1}$ is a compact operator and thus the trajectory $\lbrace U_{0}(t)\psi,t\in\bR\rbrace$ is relatively compact iff  $a_{c}$ and $a_{v}$ are bounded. If all trajectories are relatively compact one knows, see \cite{ev},  that $\sigma$ is pure point.

$E$ being $T$ periodic  $a_{c}(T)=0$ implies that $a_{c}(\cdot)$ is a bounded function.  If  $T\in2\pi\bN$ then $R(\cdot)E^{\perp}(\cdot)$ is $T$ periodic and so $a_{v}(\cdot)$ is bounded if $a_{v}(T)=0$. 
Remark that  if both  vanish then $U_0(T)$ is just multiplication by a phase factor.

If $T\notin2\pi\bN$ then the product
$R(\cdot)E^{\perp}(\cdot)$ is an almost periodic function whose associated Fourier exponents are of the form
$\lambda_{mn}=(n+m\frac{2\pi}{T})$ with $(m,n)\in\bZ\times\{-1,1\}$. Now $\inf_{\bZ\times\{-1,1\}}\vert \lambda_{mn}\vert>0$ which
implies boundedness of $a_{v}(\cdot)$, see \cite{f}, Theorem 4.12.
 \ep

\section{Compactness and the proof of Theorems \ref{thm:main} and \ref{coro:misc}}\label{sec:compactness}
In order to go from the unperturbed to the full propagator, we shall prove the following dynamical compactness result which is our main technical contribution.

\begin{theorem}\label{thm:compactness}
Let  $T_{1}<T_{2}$ be real, let $E\in C^{0}(\bR;\bR^{2})$ and $V\in C^{0}(\bR; S(1))$  Denote $U$ the propagator defined by $H(\cdot)$ in Theorem \ref{thm:propagator}.
\begin{enumerate}
\item For $f\in C^{0}(\lbrack T_{1},T_{2}\rbrack\times\bR^{2},\bR)$ and $f(t,q)\to_{\vert q\vert\to\infty}0$ uniformly in $t\in\lbrack T_1,T_2\rbrack$, it holds for the multiplication operator by $f$
\[\int_{T_1}^{T_2} U^{\ast}(t)f(t) U(t) dt \hbox{ \em \it is compact}.\]
\item For $f\in C^\infty(\lbrack T_{1}, T_{2}\rbrack\times\bR^2,\bR)$ such that for an $\varepsilon>0$ and all $\alpha \in \bN_0^2$, \\$\sup_{t}\sup_{q\in\bR^2} \vert \langle q\rangle^{(1+\varepsilon)} \partial_q^\alpha f(t,q)\vert <\infty$ we have
\[\int_{T_1}^{T_2} U_{0}^{\ast}(t)f(t) U_{0}(t) dt \hbox{ \em\it belongs to the } \alpha \hbox{\em\it-th Schatten class for } \alpha>4.\] 
\end{enumerate}
\end{theorem}

The proof is built on the following lemma

\begin{lemma}\label{lemma:sojourntime}
$T_{1}<T_{2}$ be real, let $E\in C^{0}(\bR;\bR^{2})$  and $x(t,q,p)$ defined in equation (\ref{eq:xofqp}).

\item If $f\in C^\infty(\lbrack T_{1}, T_{2}\rbrack\times\bR^2,\bR)$ such that for an $\varepsilon>0$ and all $\alpha \in \bN_0^2$, \\$\sup_{t}\sup_{q\in\bR^2} \vert \langle q\rangle^{(1+\varepsilon)} \partial_q^\alpha f(t,q)\vert <\infty$, then it holds for
$$f_{av}(q,p):=\int_{T_1}^{T_2} f(t, x(t,q,p))dt :$$

$$\sup_{(q,p)\in\bR^4}\vert\left(1+q^{2}+p^{2}\right)^{1/2}\partial_q^\alpha\partial_p^\beta f_{av}(q,p)\vert<\infty\qquad   \forall \alpha,\beta\in\bN^2_0; $$
\end{lemma}

{\bf Proof} of Lemma \ref{lemma:sojourntime}. \\

Let $h_{La},h_{0}$ be the hamiltonian symbols associated to the partial differential operators $H_{La},H_{0}(\cdot)$  
\begin{equation*}h_{La}(q,p):=\frac{1}{2}\left(p-\frac{q^{\perp}}{2}\right)^{2}, \quad
h_{0}(t,q,p):=h_{La}(q,p)-\langle E(t),q\rangle.\end{equation*}
Denote $\Phi_{h_{0}}$ {the hamiltonian flow generated by} $h_{0}$.  $x(t,q,p)  $ is the projection of $\Phi_{h_{0}}(t,q,p)$ to the q-component, i.e. the configuration space part of the trajectory, originating in $q,p$.

\[x(t,q,p)=\chi_{La}(t,q,p)+{\bf b}(t)\]
for a function ${\bf b}\in C^{0}(\bR,\bR^{2})$ and $\chi_{La}(t,q,p)$ the projection to the $q$-component of the trajectory generated by $h_{La}$.
$\chi_{La}(t,q,p)$ is linear in $q,p$ and, with the symbols of the center and velocity operators
\[v(q,p):= p-\frac{q^{\perp}}{2}, \qquad c(q,p):=q-v^{\perp},\]
one has
\[\chi_{La}(t,q,p)=c(q,p)+R(-t) v^{\perp}(q,p)\]
describing the cyclotron orbit with center $c$ and radius $\vert v\vert$. 

\medskip
Denote $z=(q,p)\in\bR^2\times\bR^2$, $\langle z\rangle:=\left(1+q^{2}+p^{2}\right)^{1/2}$.

${\bf b}$ being bounded on $\lbrack T_{1},T_{2}\rbrack$ the boundedness of $\bra q\ket^{1+\varepsilon}f(t,q)$ implies
\begin{equation}
\left\vert f\left(t, x(t,z)\right)\right\vert \le \frac{cte}{
\left(1+(\chi_{La}(t,z))^{2}\right)^{\frac{1+\varepsilon}{2}}
}\label{eq:landauflow}
\end{equation}

for a $cte>0$,  $t\in\lbrack T_{1}, T_{2}\rbrack$ and all $z$.

Denote $\widehat{z}(q,p):=\left(c(q,p),v^{\perp}(q,p)\right)\in\bR^2\times\bR^2$ and remark that
\[\frac{1}{\sqrt{2}}\vert z\vert \le \vert\widehat{z}\vert \le \sqrt{2} \vert z\vert.\]

Because of the time periodicity of $\chi_{La}$ we may assume $\lbrack T_{1},T_{2}\rbrack\subset\lbrack0, 2\pi\rbrack$. In order to estimate the decay of $\int_{T_{1}}^{T_{2}}\vert f(\chi(t,z))\vert\ dt$ we separate  the time events ``orbit is close to the origin'' and ``orbit is far from the origin'' as follows:

For a fixed number $d\in\left(0,\frac{1}{\sqrt{2}}\right)$ define for $z\in\bR^{4}$
\begin{eqnarray*}
I_{>}(z):=\lbrace t\in\lbrack0,2\pi\rbrack, \vert\chi_{La}(t,z)\vert\ge d\vert z\vert\rbrace,\quad I_{<}(z):=\lbrack0,2\pi\rbrack\setminus I_{>}
\end{eqnarray*}
and estimate the contributions to $f_{av}(z)$ of these sets.
By  (\ref{eq:landauflow}):
\[\sup_{z}\langle z\rangle\int_{\lbrack T_{1},T_{2}\rbrack\bigcap I_{>}(z)}\vert f(t, x(t,z))\vert\ dt\le\sup_{z}\frac{\langle z\rangle cte \vert T_{2}-T_{1}\vert}{\left(1+(d\vert z\vert)^{2}\right)^{\frac{1+\varepsilon}{2}}}<\infty.\]

Now for $z$ fixed let $t_{0}\in\lbrack0,2\pi\rbrack$ be the point for which the distance to the origin $t\mapsto\vert \chi_{La}(t,z)\vert$ is minimal.\\ If $I_{<}(z)\neq\emptyset$ then $t_{0}\in I_{<}(z)$. It holds
\[\partial_{t}^{2}\vert\chi_{La}(t,z)\vert^{2}=-2\langle c, R(-t)v^{\perp}\rangle\]
and thus for $t\in I_{<}(z):$
\[\vert\chi_{La}(t,z)\vert^{2}=c^{2}+v^{2}+2\langle c,R(-t) v^{\perp}\rangle\le 2d^{2}\vert\widehat{z}\vert^{2}=:{\widehat d}^{2}\vert\widehat{z}\vert^{2}, \hbox{ \em\it which implies}\]
\[\partial_{t}^{2}\vert\chi_{La}(t,z)\vert^{2}\ge(1-{\widehat d}^{2})\vert\widehat{z}\vert^{2}. \hbox{ \em\it It follows that for a\ }\tau_{t}\in\lbrack t_{0},t\rbrack\]
\[\vert\chi_{La}(t,z)\vert^{2}=\vert\chi_{La}(t_{0},z)\vert^{2}+\partial_{t}^{2}\vert\chi_{La}(\tau_{t})\vert^{2}\frac{(t-t_{0})^{2}}{2}\ge\frac{(1-{\widehat d}^{2})}{2}\vert\widehat{z}\vert^{2}(t-t_{0})^{2} \hbox{ \em\it and}\]
\[\sup_{z}\langle z\rangle\int_{\lbrack T_{1},T_{2}\rbrack\bigcap I_{<}(z)}\vert f(t, x(t,z))\vert\ dt\le\sup_{z} \int_{T_{1}}^{T_{2}}\frac{cte \langle z\rangle}{\left(1+\left(\frac{1-{\widehat d}^{2}}{2}\right)\vert\widehat{z}\vert^{2}(t-t_{0})^{2}\right)^{\frac{1+\varepsilon}{2}}}\ dt<\infty\]

and the  claim is proven for $f$. Analogously the assertion that the derivatives decay follows  from the assumptions on the derivatives of $f$ and the affine character of $z\mapsto x(t,z)$.
\ep

\medskip

\begin{remark}
Intuitively Lemma \ref{lemma:sojourntime} means that the sojourn time of a classical particle in the "support"  of $f$ decays as $1/ \left(1+q^{2}+p^{2}\right)^{1/2}$ in the initial data because the center and the radius of the cyclotron orbit grow linearly in these data.
\end{remark}
\bigskip

{\bf Proof} of Theorem \ref{thm:compactness}. \\

We use pseudodifferential Weyl Calculus, see \cite{fo,zw} and references therein. We first prove  the second assertion.

 By Egorov's Theorem $U_{0}^{\ast}(t)f(t) U_{0}(t)$ (respectively  its integral) is the Weyl quantization of the symbol $z\mapsto f(x(t,z))$ (respectively $\int_{T_1}^{T_2} f(x(t,z))\ dt$).  Lemma \ref{lemma:sojourntime} means that
$$\int_{T_1}^{T_2} f(t, x(t,.))\ dt\in S(m,g),$$
the H\"ormander class with respect to the euclidean metric
$$g=\sum_i dq_i^2+dp_i^2$$
and weight function
$$m(z)=\langle z\rangle=(1+q^2+p^2)^{\frac{1}{2}}.$$
$m$ and $\int_{T_1}^{T_2} f(t, x(t,.))dt$ are both in $L^\alpha(\bR^4)$ for $\alpha>4$. Thus the second claim follows from Theorem 2.1 of \cite{bt}, see also their Proposition 4.2. 

\medskip 
Now $C_{0}^{\infty}(\bR)$ is uniformly dense in the continuous functions vanishing at infinity and  the space of compact operators in the bounded operators is norm closed.
Thus we  conclude that

\[\int_{T_1}^{T_2} U_{0}^{\ast}(t)f(t) U_{0}(t) dt \]
is compact
 for $f\in C^{0}(\lbrack T_{1},T_{2}\rbrack\times\bR^{2},\bR)$ and $f(t,q)\to_{\vert q\vert\to\infty}0$ uniformly in $t$. 
 Now suppose that $B(t)\quad (t\in\lbrack0,T\rbrack)$ is a norm continuous family of operators. We prove:
\[\int_{s}^{t}U_{0}^{\ast}(\tau)B(\tau)U_{0}(\tau)\ d\tau \hbox{ \em\it compact } (\forall s,t\in\lbrack0,T\rbrack)\Longrightarrow \int_{0}^{T}U^{\ast}(\tau)B(\tau)U(\tau)\ d\tau \hbox{ \em\it is compact}.\]

We follow an argument of \cite{ev} :  denote $\Omega(t):=U^{\ast}_{0}(t)U(t)$; $\lbrack0,T\rbrack$ is the disjoint union of  $I_{j}=\frac{T}{N}\lbrack j,j+1), \quad\{j\in0,\cdots,N-1\}$.
Note that
\begin{equation*}\label{eq:diffpro}
\Vert \Omega(t)-\Omega(s)\Vert\le\vert t-s\vert \Vert V\Vert.\end{equation*}
Now use that  for $t,s\in I_{j}$, $\tau\in(t,s)$ it holds
$\int_{s}^{t}\Vert B\Vert=\cO\left(\frac{1}{N}\right)$ and $\Vert\Omega(\tau)-\Omega(s)\Vert=\cO\left(\frac{1}{N}\right)$ thus
\[\int_{s}^{t}U^{\ast}BU(\tau)\ d\tau=\Omega^{\ast}(s)\left(\int_{s}^{t}U_{0}^{\ast}BU_{0}(\tau)\ d\tau\right) \Omega(s)+\cO\left(\frac{1}{N^{2}}\right).\]
It follows
\[\left\Vert\int_{0}^{T}U^{\ast}BU-\sum_{j=0}^{N-1}\Omega^{\ast}({jT}/{N}){\left(\int_{I_{j}}U_{0}^{\ast}BU_{0}\right)}\Omega({jT}/{N})\right\Vert=\cO\left(\frac{1}{N}\right).\]
Since $\int_{I_{j}}U_{0}^{\ast}BU_{0}$ is compact $\forall j$, we conclude that $\int_{0}^{T}U^{\ast}BU$ is compact as the space of compact operators is norm closed. 

In particular $\int_{0}^{T}U^{\ast}(t)K(t)U(t)\ dt$ is compact so the theorem is proven.
\ep

We now prove the Mourre estimate needed to apply Theorem \ref{thm:abcf}.

\begin{theorem}\label{thm:fulldynamics}	
Assume {\bf (A)}. Denote $U$ the propagator defined by $H(\cdot)$ in Theorem \ref{thm:propagator} and the quantities $a_{c}, A_{c},a_{v},A_{v}$ defined in  (\ref{eq:acav},\ref{eq:aoperators}).
\begin{enumerate}
\item Suppose $\int_{0}^{T}E\neq 0$ and $\bra \int_{0}^{T}E^{\perp},\nabla V(t,q)\ket\to_{\vert q\vert\to\infty}0$ uniformly in $t$. Then there exists a compact operator $\cC$ on $L^{2}(\bR^{2})$ such that
\begin{eqnarray*}
(U^{\ast}(T)A_{c}U^{\ast}(T)-A_{c})=\vert a_{c}(T)\vert^{2}\bI + \cC,\\
 \left\lbrack \left(U^{\ast}(T)A_{c}U(T)-A_{c}\right) ,A_{c}\right\rbrack \hbox{ \em\it is bounded. }
\end{eqnarray*}


\item Suppose $T\in2\pi\bN$ and  $\vert \nabla V(t,q)\vert\to_{\vert q\vert\to\infty}0$ uniformly in $t$. If $\int_{0}^{T}R(s)E(s)\ ds\neq0$  then there exists a compact operator $\cC$ on $L^{2}(\bR^{2})$ such that
\begin{eqnarray*}
(U^{\ast}(T)A_{v}U^{\ast}(T)-A_{v})=\vert a_{v}(T)\vert^{2}\bI + \cC,\\
 \left\lbrack \left(U^{\ast}(T)A_{v}U(T)-A_{v}\right) ,A_{v}\right\rbrack \hbox{ \em\it is bounded. }
\end{eqnarray*}
\end{enumerate}
\end{theorem}

{\bf Proof.} Throughout  this proof the symbols $a, A$ without subscript denote one of the $a_{c}, A_{c}$ or $a_{v}, A_{v}$  defined in (\ref{eq:acav},\ref{eq:aoperators}).To prove the results concerning the first commutator 
we first determine a continuous family of bounded  operators $K$ such that  
\[(U^{\ast}(T)AU(T)-A)-\vert a\vert^{2}=-i\int_{0}^{T}U^{\ast}(t)K(t)U(t)dt,\]
and such that $\int_{0}^{T}U^{\ast}(t)K(t)U(t)dt$ is compact, then we shall conclude by applying Theorem \ref{thm:abcf}.

Each of the following computations is to be understood  first on $\cS(\bR^{2})$ then by extension to $L^{2}(\bR^{2})$; recall Theorem \ref{thm:propagator} and in particular that all the involved operators leave $\cS(\bR^2)$ invariant.

Suppose $T\in2\pi\bN$. By (\ref{eq:Mourrecfree}, \ref{eq:Mourrevfree}) : $\left(U^{\ast}_{0}(T)AU_{0}(T)-A\right)=\vert a\vert^{2}$. It follows
 \[U^{\ast}(T)AU(T)-\vert a\vert^{2}-A=U^{\ast}U_{0}(T)AU_{0}^{\ast}U(T)-A\]
\begin{equation}\label{eq:thirteen}=-i \int_{0}^{T}U^{\ast}(t)\underbrace{\left\lbrack U_{0}(t)AU_{0}^{\ast}(t),V\right\rbrack}_{=: K(t)} U(t)\ dt.\end{equation}

Remark that for the case $A_c, a_c$ (\ref{eq:thirteen}) holds for any $T$. 

From the explicit free evolution of observables,  Corollary \ref{coro:freepropagator}, one gets
\[K_{c}(t)=\left\lbrack U_{0}(t)A_{c}U_{0}^{\ast}(t),V(t)\right\rbrack=\left\lbrack \bra a_{c}(T),-D^{\perp}\ket,V(t)\right\rbrack=i\bra a_{c}(T),\nabla V^{\perp}(t)\ket\]
and
\[K_{v}(t)=\left\lbrack \bra a_{v}(T),R(-t)D^{\perp}\ket,V(t)\right\rbrack=-i\bra R(t)a_{v}(T),\nabla V^{\perp}(t)\ket.\]

By the decay assumption  on  $\nabla V$ Theorem \ref{thm:compactness} is applicable which in both cases implies the compactness of $\int_{0}^{T}U^{\ast}(t)K(t)U(t)dt$. Note that (\ref{eq:thirteen})  implies in particular that $(\Omega^\ast(T)A\Omega(T)-A)$, and hence $\lbrack A, \Omega(T)\rbrack$, is bounded.

\bigskip

Concerning the double commutator observe :
\[\left\lbrack \left(U^{\ast}(T)AU(T)-A\right), A\right\rbrack=_{(\ref{eq:thirteen})}\left\lbrack\left(\Omega^{\ast}(T)A\Omega(T)-A\right),A\right\rbrack=\]
\[\left\lbrack \Omega^{\ast}(T)\left\lbrack A,\Omega(T)\right\rbrack, A\right\rbrack=\Omega^\ast(T)\left\lbrack\left\lbrack A,\Omega(T)\right\rbrack, A\right\rbrack+\Omega^\ast(T)\left\lbrack A, \Omega(T)\right\rbrack\Omega^\ast(T)\left\lbrack A,\Omega(T)\right\rbrack.\]
Thus it is sufficient to prove that  $\left\lbrack A, \Omega(t)\right\rbrack$ and $\left\lbrack A, \left\lbrack A, \Omega(t)\right\rbrack\right\rbrack$ admit bounded extensions to infer that $\left\lbrack \left(U^{\ast}(T)AU(T)-A\right), A\right\rbrack$ can be extended to a bounded operator.
By Theorem \ref{thm:propagator}.\ref{4} $\Omega$ is the solution of 
\[i\partial_{t}\Omega(t)=V(t,x(t,x,D)\Omega(t), \quad \Omega(0)=\bI,\]
thus denoting $L(t):=-V(t,x(t,x,D)$  the operator triple
\[OT(t):=\left(\Omega(t), \left\lbrack A, \Omega(t)\right\rbrack, \left\lbrack A, \left\lbrack A, \Omega(t)\right\rbrack\right\rbrack\right)\]
 is a fixed point of the system
\begin{eqnarray}
&&\Omega(t)=\bI+i\int_{0}^{t}L \Omega\nonumber\\
&&\left\lbrack A,\Omega(t)\right\rbrack=i\int_{0}^{t}\left(\left\lbrack A,L\right\rbrack\Omega+L\left\lbrack A,\Omega\right\rbrack\right)\label{eq:fourteen}\\
&&\left\lbrack A,\left\lbrack A,\Omega(t)\right\rbrack\right\rbrack=i\int_{0}^{t}\left(\left\lbrack A, \left\lbrack A, L\right\rbrack\right\rbrack\Omega+2\left\lbrack A,L\right\rbrack\left\lbrack A,\Omega\right\rbrack+L\left\lbrack A, \left\lbrack A,\Omega\right\rbrack\right\rbrack\right).\nonumber
\end{eqnarray}
The right hand side as applied to $OT$ is a Volterra operator in the Banach space $\bK^{3}$ where $\bK:=C^{0}\left(\lbrack0,T\rbrack,\bB\left(L^{2}(\bR^{2})\right)\right)$ with the norm $\sup_{t}\Vert\cdot\Vert$
provided that $L, \left\lbrack A,L\right\rbrack$ and $\left\lbrack A,\left\lbrack A,L\right\rbrack\right\rbrack$ belong to $\bK$ which we will show now.

$V$ is bounded thus $L\in\bK$. 
$\left\lbrack A,L\right\rbrack=-U^{\ast}_{0}\left\lbrack U_{0}AU^{\ast}_{0},V\right\rbrack U_{0}=_{(\ref{eq:thirteen})}-U^{\ast}_{0}K(t)U_{0}$
and it follows that $\left\lbrack A,L\right\rbrack\in\bK$. Finally
\[\left\lbrack A,\left\lbrack A,L\right\rbrack\right\rbrack=-U^{\ast}_{0} \left\lbrack U_{0}AU_{0}^{\ast},K\right\rbrack U_{0}.\]
Using the explicit expressions for $K$ we get
\[\left\lbrack U_{0}A_{c}U_{0}^{\ast},K_{c}\right\rbrack=\left\lbrack\bra a_{c}(T), D^{\perp}\ket, \left\lbrack\bra a_{c}(T), D^{\perp}\ket,V\right\rbrack\right\rbrack=\bra a_{c}(T)^{\perp},Hess(V) a_{c}(T)^{\perp}\ket\]

\begin{eqnarray*}
\left\lbrack U_{0}A_{v}U_{0}^{\ast},K_{v}\right\rbrack(t)&=&\left\lbrack\bra R(t)a_{v}(T), D^{\perp}\ket, \left\lbrack\bra R(t)a_{v}(T), D^{\perp}\ket,V(t)\right\rbrack\right\rbrack\\
&=&\bra R(t)a_{v}(T)^{\perp},Hess(V(t)) R(t)a_{v}(T)^{\perp}\ket.
\end{eqnarray*}

Thus from our assumptions of boundedness of the second derivatives of $V$ it follows that $\left\lbrack A,\left\lbrack A,L\right\rbrack\right\rbrack\in\bK$. We conclude that the fixed point $OT\in\bK^3$ so the double commutators admit bounded extensions. Theorem \ref{thm:abcf} is applicable and the proof of Theorem \ref{thm:fulldynamics} is completed.\ep

\bigskip

We now have all elements to provide the proofs of  Theorems \ref{thm:main} and \ref{coro:misc}

\medskip

{\bf Proof.} of Theorem \ref{thm:main}

The spectral results, i.e. point 1. for the case $T\in2\pi\bN$ and point 1.(a) for $T\notin2\pi\bN$, follow from Theorem \ref{thm:abcf} and Theorem \ref{thm:fulldynamics} because \\ $\vert a_c(T)\vert+\vert a_v(T)\vert\neq0$. 

To prove the results on $v_{asy}$, recall equations (\ref{eq:qoftfree},\ref{eq:coftfree},\ref{eq:voftfree}) for the free evolution of observables. 

Let  $\psi\in\cS(\bR^2)$. As in equation (\ref{eq:thirteen}) one calculates:
\begin{eqnarray*}
&&U^\ast(t)v^\perp U(t)\psi=U^\ast U_0(t)\left(R(-t)v^\perp+\int_0^tR(s-t)E^\perp(s)ds\right)U_0^\ast U(t)\psi\\
&&=\int_0^t R(s-t)E^\perp(s)ds\psi+R(-t)\left(v^\perp\psi+i\int_0^t U^\ast(s)\left\lbrack R(s) D^\perp,V(s)\right\rbrack U(s)\psi\ ds\right)\\
&&=\int_0^t R(s-t)E^\perp(s)ds\psi+R(-t)v^\perp\psi+\int_0^t U^\ast(s)R(s)\nabla V^\perp(s) U(s)\psi\ ds
\end{eqnarray*}
and 
\[U^\ast(t) c U(t)\psi=c\psi-\int_0^t E^\perp\psi+\int_0^t U^\ast(s)\left(-\nabla V^\perp(s)\right)U(s)\psi\ ds,\]
\begin{equation}
\label{eq:qoft}U^\ast(t) q U(t)\psi=U^\ast_0(t) q U_0(t)\psi+\underbrace{\int_0^t U^\ast(s)\left(R(s-t)-\bI\right)\nabla V^\perp(s) U(s)\psi\ ds.}_{=: \cC_q(t)\psi}
\end{equation}

It follows from $U(t+T,t_0+T)=U(t,t_0)$ that
\[\cC_q(nT)=\sum_{j=0}^{n-1}U^\ast(jT)\cC_q(T)U(jT)\quad(n\in\bN).\]
By the decay hypothesis on $\nabla V$ and  Theorem \ref{thm:compactness} we have that $\cC_q(T)$ is compact.
Now

\begin{eqnarray}
&&\frac{1}{nT}\left(U^\ast(nT)qU(nT)-q\right)\psi=
\frac{1}{nT}\left(R(-nT)-\bI\right)v^\perp\psi\label{eq:3terms}\\
&&+
\frac{1}{nT}\int_0^{nT}(R(s-nT)-\bI)E^\perp(s)\ ds\psi+
\frac{1}{nT}\sum_{j=0}^{n-1}U^\ast(jT)\cC_q(T)U(jT)\psi.\nonumber\end{eqnarray}

The first term on the right hand side of (\ref{eq:3terms}) converges to zero when $n$ goes to infinity. For the third term it holds by Wiener's theorem, see Chapter 5.4 in \cite{cfks}:

\[s-\lim_{n\to\infty}\frac{1}{nT}\sum_{j=0}^{n-1}U^\ast(jT)\cC_q(T)U(jT)=\sum_{\lambda\in\sigma_d}P_\lambda \cC_q(T)P_\lambda\ ,\]
where $\sigma_d$ denotes the (finite) set of eigenvalues of $U(T)$ and $P_\lambda$ the eigenprojection associated to $\lambda\in\sigma_d$.

Now

\[\frac{1}{nT}\int_0^{nT}(R(s-nT)-\bI)E^\perp(s)\ ds\to_{n\to\infty}\left\{\begin{array}{ll}
 \frac{1}{T}\int_0^T\left(R(s)-\bI\right)E^\perp(s)\ ds &T\in2\pi\bN    \\
- \frac{1}{T}\int_0^T E^\perp(s)\ ds &T\notin2\pi\bN      
\end{array}\right.
\]

because $\lim_{\tau\to\infty}\frac{1}{\tau}\int_0^\tau RE^\perp=0$ if $T\notin2\pi\bN$ as remarked in the proof of Proposition \ref{propo:freeMourre}. 

Using the symbol $q(nT):=U^\ast(nT)qU(nT)$ we have shown that for $\psi\in\cS(\bR^2)$:
\[\lim_{n\to\infty}\frac{1}{nT}\left(q(nT)-q\right)\psi=\lim_{n\to\infty}\frac{1}{nT}q(nT)\psi=v_{asy}\psi\]
with the bounded operator
\[v_{asy}:=\left\{\begin{array}{ll}
 \frac{1}{T}\int_0^T\left(R(s)-\bI\right)E^\perp(s)\ ds +\sum_{\lambda\in\sigma_d}P_\lambda \cC_q(T)P_\lambda &T\in2\pi\bN    \\
- \frac{1}{T}\int_0^T E^\perp(s)\ ds + \sum_{\lambda\in\sigma_d}P_\lambda \cC_q(T)P_\lambda&T\notin2\pi\bN.    
\end{array}\right.
\]
In order to show the claimed result on $v_{asy}$ we shall now prove that a virial theorem holds, meaning here:   $P_\lambda v_{asy} P_\lambda=0\quad\forall\lambda\in\sigma_d$. This is not trivial because $\frac{1}{nT}\left(q(nT)-q\right)$ is unbounded in general.

Let $0\neq x\in\bR^2$. Then for $\psi\in \cS(\bR^2),\quad\varphi=\left(i+\left\langle x,v_{asy}\right\rangle\right)\psi$
\begin{eqnarray*}
&\left(\left(i+\left\langle x,{q(nT)}/ {nT}\right\rangle\right)^{-1}-\left(i+\left\langle x,v_{asy}\right\rangle\right)^{-1}\right)\varphi=\\
&\left(i+\left\langle x,{q(nT)}/ {nT}\right\rangle\right)^{-1}\left\langle x,\left(v_{asy}-{q(nT)}/ {nT}\right)\right\rangle\psi \to_{n\to\infty}0.
\end{eqnarray*}
Now $v_{asy}$ is bounded implying that  $\left(i+\left\langle x,v_{asy}\right\rangle\right)\cS(\bR^2)$ is dense in $L^2(\bR^2,dq)$, $\left(i+\left\langle x,v_{asy}\right\rangle\right)^{-1}$ and $\left(i+\left\langle x,{q(nT)}/ {nT}\right\rangle\right)^{-1}$ are bounded by $1$ so it follows that 
\[s-\lim_{n\to\infty}\left(i+\left\langle x,{q(nT)}/ {nT}\right\rangle\right)^{-1}=\left(i+\left\langle x,v_{asy}\right\rangle\right)^{-1}.\]
On the other hand
\[P_\lambda \left(i+\left\langle x,{q(nT)}/ {nT}\right\rangle\right)^{-1} P_\lambda=P_\lambda \left(i+\left\langle x,{q}/ {nT}\right\rangle\right)^{-1}P_\lambda\to_{n\to\infty}\]\[0=
P_\lambda\left(i+\left\langle x,v_{asy}\right\rangle\right)^{-1}P_\lambda.\]
This is true for all $x$ so it follows that $P_\lambda v_{asy} P_\lambda=0$ and we have finished the proof that
\[v_{asy}=\left\{\begin{array}{ll}
 \frac{1}{T}\int_0^T\left(R(s)-\bI\right)E^\perp(s)\ ds\  P_{ac}(U(T)) &T\in2\pi\bN    \\
- \frac{1}{T}\int_0^T E^\perp(s)\ ds\  P_{ac}(U(T)) &T\notin2\pi\bN.    
\end{array}\right.
\]
\medskip

It remains to prove the energy growth in the case $T\in2\pi\bN$.
By the same argument as before we have for the asymptotic acceleration:
\[s-\lim_{n\to\infty}\frac{1}{nT}\left(\underbrace{U^\ast(nT)v^\perp U(nT)}_{=:v^\perp(nT)}-v^\perp\right)=s-\lim_{n\to\infty}\frac{1}{nT}v^\perp(nT)=\frac{1}{T}\int_0^T RE^\perp P_{ac}(U(T)).\]

Now for $\psi\in\cS(\bR^2)$:
\begin{eqnarray*}
&&\frac{1}{(nT)^2}\left\langle U^\ast(nT)\psi, H_{La} U(nT)\psi\right\rangle =\frac{1}{2}\sum_{j=1}^2\Vert\frac{1}{nT}v_j^\perp(nT)\psi\Vert^2\\
&&\to_{n\to\infty}\frac{1}{2}\left\vert\frac{1}{T}\int_0^T RE^\perp\right\vert^2 \Vert P_{ac}(U(T))\psi\Vert^2
\end{eqnarray*}
\ep

\bigskip

{\bf Proof.} of Theorem \ref{coro:misc}

1) From the proof of Theorem \ref{thm:fulldynamics} we have 

\[(U^{\ast}(T)A_{c}U^{\ast}(T)-A_{c})=\vert a_{c}(T)\vert^{2}\bI 
+  \int_{0}^{T}U^{\ast}(t)\bra a_{c}(T),\nabla V^{\perp}(t)\ket U(t)\ dt.\]

\[(U^{\ast}(T)A_{v}U^{\ast}(T)-A_{v})=\vert a_{v}(T)\vert^{2}\bI 
-  \int_{0}^{T}U^{\ast}(t)\bra R(t)a_{v}(T),\nabla V^{\perp}(t)\ket U(t)\ dt.\]

The assumption of smallness of $\nabla V$  implies that one of  right hand sides is a positive operator and the result follows from Theorem \ref{thm:abcf}

2)
 Theorem \ref{thm:compactness}  and the assumed decay of $V$ imply compactness of
\[\int_0^T U_0^\ast(t)V(t)U_0(t)\ dt\]
which as in the proof of Theorem \ref{thm:compactness}  implies compactness of 
\[\int_0^T U_0^\ast(t)V(t)U(t)\ dt=U_0^\ast(T)U(T)-\bI.\]
Thus $U(T)-U_0(T)$ is compact and the essential spectra of $U(T)$ and $U_0(T)$ coincide. By formula (\ref{eq:freepropagator}) $U_0(T)=e^{-i\varphi(T)}e^{-iT H_{La}}$  so its spectrum is the  set of points $\left(e^{-i\varphi(T)}e^{-i(n+\frac{1}{2}) T}\right)_{n\in\bN_0}$ which is discrete as $T\in2\pi\bQ$ which implies that the spectrum  of $U(T)$ is pure point  and  Theorem \ref{coro:misc} is proven.
\ep

Remark that this argument  on stability of the pure point spectrum for resonant perturbations was  applied in \cite{ade} to the electric AC Stark effect.

\bigskip
{\bf Acknowledgements}

We  gratefully acknowledge  support from the
grants Fondecyt Grant 1120786;  CONICYT PIA-ACT1112; ECOS-CONICYT C10E01. OB thanks CPT, JA and CM thank Facultad de Matem\'aticas of PUC for hospitality.

\end{document}